\documentclass[
    reprint,
    amsfonts,
    amssymb,
    amsmath, 
    aps,
    prd,
    nofootinbib
    ]{revtex4-1}

\usepackage[latin1]{inputenc}
\usepackage[T1]{fontenc}
\usepackage[english]{babel}
\usepackage{booktabs}
\usepackage{tabularx}
\usepackage[pdftex,table]{xcolor}
\usepackage{lieart}
\usepackage[vcentermath]{youngtab}

\usepackage[pdftex,bookmarksopen]{hyperref}
\hypersetup{
  pdftitle={Grand Unification and Exotic Fermions}, 
  pdfauthor={Robert Feger}, 
  pdfsubject={}, 
  pdfcreator={pdfTeX},
  pdfproducer={Robert Feger}, 
  pdfkeywords={},
}

\begin{document}

\title{Grand Unification and Exotic Fermions}

\author{Robert P. Feger}
\email{robert.feger@physik.uni-wuerzburg.de}
\affiliation{Universit\"at W\"urzburg, Institut f\"ur Theoretische Physik und Astrophysik, Emil-Hilb-Weg 22, 97074 W\"urzburg, Germany}

\author{Thomas W. Kephart}
\email{thomas.w.kephart@vanderbilt.edu}
\affiliation{Department of Physics and Astronomy, Vanderbilt University, Nashville, Tennessee 37235, USA}

\date{\today}

\begin{abstract}
We exploit the recently developed software package LieART to show that \SU{N} 
grand unified theories with chiral fermions in mixed tensor irreducible 
representations can lead to standard model chiral fermions without additional 
light exotic chiral fermions, i.e., only standard model fermions are light in 
these models. Results are tabulated which may be of use to model builders in the 
future. An \SU{6} toy model is given and model searches are discussed.

\end{abstract}
\maketitle

\section{Introduction}
\label{sec:Introduction}
In the past, building grand unified theories (GUTs) with \SU{N} gauge groups has 
nearly always been carried out using fermions in totally antisymmetric tensor 
irreducible representations (irreps). Choosing a chiral anomaly free set of 
these \SU{N} irreps guarantees all fermions will continue to be anomaly free and 
in totally antisymmetric irreps when decomposed into regular \SU{N'} subgroups 
with $N' < N$. We will typically choose $N' = 5$. Hence, under the decomposition
\[\SU{N}\rightarrow \SU{5}\]
we have
\begin{multline}
 {asym~anomaly~free~SU(N)~irreps}\\
 \rightarrow
 n(\irrepbar{5} +\irrep{10}) + \bar{n}(\irrep{5}+\irrepbar{10})+ \text{singlets} 
\end{multline}
so that $n_F=n-\bar{n}$ gives the number of families. There are only a few cases 
of studies of \SU{N} models where other than totally antisymmetric irreps have 
been used. For example, single complex anomaly free irreps of \SU{N} that 
contain chiral fermions have been searched for \cite{Eichten:1982pn}, and models 
with fermions in \irrep{6}s and \irrep{8}s of \SU{3} color have been studied 
\cite{Frampton:1987dn}. Here we ask if there are \SU{N} models that start with 
fermions in complex mixed tensor irreps that lead to models with only standard 
model (SM) chiral fermions being light. The simplest way to explore such \SU{N} 
models is to require that the only chiral fermions at the \SU{5} level are in 
standard $(\irrepbar{5}+\irrep{10})$s families which then lead to 
$\SU{3}\times\SU{2}\times\U{1}$ standard model families 
\begin{equation}
\irrepbar{5} + \irrep{10} \rightarrow (\irrep{3},\irrep{2})_\frac{1}{6} +(\irrepbar{3},\irrep{1})_\frac{1}{3}+(\irrepbar{3},\irrep{1})_{{-}\frac{2}{3}}+ (\irrep{1}, \irrep{2})_{{-}\frac{1}{2}}+(\irrep{1},\irrep{1})_1
\end{equation}\\
However, GUT models~\cite{Kim:1979vm} and partial gauge 
unifications~\cite{Frampton:1987dn,Kephart:2001ix,Kephart:2006zd} with exotic 
fermions are not unknown. Exotics from string 
theory~\cite{Dienes:1995sq,Barger:2006fm} and 
F-Theory~\cite{Leontaris:2009wi,Callaghan:2013kaa} have also been considered.

\section{\boldmath General \SU{N} Models}
Let us focus on the decomposition $\SU{N}\rightarrow \SU{5}$. A totally 
antisymmetric \SU{N} tensor irrep corresponds to single column Young tableau. 
All \SU{N} single column tableaux decompose to a single column \SU{5} tableau 
under the regular embedding. In addition, if a set of \SU{N} irreps is anomaly 
free, then so is the set of \SU{5} irreps they decompose into. These two facts 
are the reason models can be successfully constructed in \SU{N} gauge theories 
that reduce to exotic free models at the SM level.

Now we ask if it is still possible to build chiral \SU{N} models that are both 
anomaly free and exotic free at the \SU{5} and hence the SM level if we start 
with irreps that correspond to other than single column tableaux. We will begin 
with the case of models with fermions in irreps corresponding to two-column 
tableaux. These irreps can only decompose into one and two-column tableaux of 
\SU{5}. (More generally, an $n$ column tableau of \SU{N} can decompose into $n, 
n-1,...,n-k$ column tableaux of \SU{N-k}.) Hence we would like to find a set of 
chiral anomaly free two-column \SU{N} tableaux that decompose such that the 
resulting two-column set in \SU{5} is vector-like, while at least part of the 
one column set remains chiral and anomaly free. These chiral fermions must then 
be in the form of standard $(\irrepbar{5}+\irrep{10})$s families.

In the past this type of model has been difficult to explore, but we now have a 
tool in hand that makes the work quite easy. The software package LieART%
\footnote{LieART is hosted by Hepforge, IPPP Durham. The LieART project home 
page is \href{http://lieart.hepforge.org}{http://lieart.hepforge.org} and the 
LieART Mathematica application can be freely downloaded as a tar.gz archive from 
\href{http://www.hepforge.org/downloads/lieart}{%
http://www.hepforge.org/downloads/lieart}}\cite{Feger:2012bs}, written in 
Mathematica, can be used to project combinations of multicolumn \SU{N} tableaux 
to \SU{5} efficiently and keep track of the chirality in going from \SU{N} to 
\SU{5}. Our results are displayed in the tables in the next section and other 
possible searches are discussed. An \SU{6} toy model is given in section IV 
before we conclude in section V. Checking any of these results by hand will 
clearly demonstrate the power and flexibility of LieART. 

\section{Results}
\label{sec:results}

\newcommand{\irrepstrut}{\rule[-.3\baselineskip]{0pt}{1.2\baselineskip}}
\newcommand{\irrepandtableau}[2]{\parbox[t][][c]{0.15\columnwidth}{\irrepstrut#1\\#2}}
\begin{table}
	\begin{center}
		\Yautoscale0
		\Yboxdim13pt
		\begin{tabular}{cccccccc}
			\irrepandtableau{\irrep{21}}{\yng(2)}&
			\irrepandtableau{\irrep{70}}{\yng(2,1)}&
			\irrepandtableau{\irrepbar{105}}{\yng(2,1,1)}&
			\irrepandtableau{\irrepbar{84}}{\yng(2,1,1,1)}&
			\irrepandtableau{\irrep{35}}{\yng(2,1,1,1,1)}\\
			\irrepandtableau{\irrepbar[1]{105}}{\yng(2,2)}&
			\irrepandtableau{\irrepbar{210}}{\yng(2,2,1)}&
			\irrepandtableau{\irrep{189}}{\yng(2,2,1,1)}&
			\irrepandtableau{\irrep{84}}{\yng(2,2,1,1,1)}&
			\irrepandtableau{\irrep{175}}{\yng(2,2,2)}\\
			\irrepandtableau{\irrep{210}}{\yng(2,2,2,1)}&
			\irrepandtableau{\irrep{105}}{\yng(2,2,2,1,1)}&
			\irrepandtableau{\irrep[1]{105}}{\yng(2,2,2,2)}&
			\irrepandtableau{\irrepbar{70}}{\yng(2,2,2,2,1)}&
			\irrepandtableau{\irrepbar{21}}{\yng(2,2,2,2,2)}\\
		\end{tabular}
	\end{center}
	\caption{\label{tab:twocolumnSU6} The two-column tableaux for \SU{6}. Note that the \irrep{35}, \irrep{189} and \irrep{175} are all real so will not contribute chiral fermions.}
\end{table}

Let us begin with the simplest example we have found---an \SU{6} model with only 
two-column tableaux as displayed in Table~\ref{tab:twocolumnSU6}. 

The non-conjugated, complex, two-column tableaux irreps of \SU{6} decompose to 
\SU{5} irreps as
\begin{equation}\label{eqn:TwoColumnTableauxSU6toSU5Decomposition}
	\begin{array}{l@{\;\rightarrow\;}l}
		\irrep{21}     & \irrep{1}    + \irrep{5}    +\irrep{15}              \\
		\irrep{70}     & \irrep{5}    + \irrep{10}   +\irrep{15}+\irrepbar{40}\\
		\irrep{84}     & \irrep{5}    + \irrep{10}   +\irrep{24}+\irrep{45}   \\
		\irrep{105}    & \irrep{10}   + \irrepbar{10}+\irrep{40}+\irrep{45}   \\
		\irrep[1]{105} & \irrepbar{15}+ \irrep{40}   +\irrep{50}              \\
		\irrep{210}    & \irrep{40}   + \irrep{45}   +\irrep{50}+\irrep{75}   \\
	\end{array}
\end{equation}
and the complex conjugated irreps decompose analogously. One then just has to 
find linear combinations of \SU{6} irreps with three families that are free from 
exotics at the \SU{5} level, which for \SU{6} delivers the single example
\begin{equation}\label{eqn:SU6Solution}
	6(\irrepbar{21}) + 9(\irrep{70}) + 6(\irrepbar{84}) + 9(\irrep{105}) + 3(\irrep[1]{105}) + 3(\irrepbar{210})
\end{equation}
which when decomposed into \SU{5} irreps reduces to 
\begin{align}
	\begin{split}
		3(\irrep{10}+ \irrepbar{5}) 
		& + 9( \irrep{5}+ \irrepbar{5}) 
		+ 15(\irrep{10}+\irrepbar{10}) \\
		& +  9(\irrep{15}+\irrepbar{15}) 
		+ 12(\irrep{40}+\irrepbar{40}) \\
		& +  9(\irrep{45}+\irrepbar{45}) 
		+  3(\irrep{50}+\irrepbar{50}) \\
		& +  6(\irrep{1}) 
		+  6(\irrep{24}) 
		+  3(\irrep{75}) 
	\end{split}
\end{align}
where all irreps not belonging to the three families come in conjugated pairs, thus being vector-like.

More generally we implemented an efficient determination of exotic-free 
combinations of mixed tensor irreps of \SU{N} utilizing LieART. The requirement 
of three families and no chiral exotics at the \SU{5} level leads to a system of 
linear equations which reduces the number of independent parameters being 
initially one per irrep type. To this end we introduce special multiplicities $m_i$
coding the imbalance of complex-conjugated and non-conjugated irrep pairs, i.e., 
a positive multiplicity denotes an excess of non-conjugated irreps and a 
negative multiplicity an excess of conjugated irreps. For the \SU{6} model with 
only two-column tableaux the ansatz for the determination of an exotic-free, 
three SM family model reads
\begin{equation}
	\begin{aligned}
	m_1 \irrep{21} + m_2\, \irrep{70} + m_3\, \irrep{84} + m_4\, \irrep{105} + m_5\, \irrep[1]{105} + m_6\, \irrep{210}\\
	\to  -3(\irrep{5}) + 3(\irrep{10}) + 0 (\irrep{15}) + 0(\irrep{40}) + 0 (\irrep{45}) + 0 (\irrep{50}).
	\end{aligned}
\end{equation}
Note that real irreps such as \irrep{1}, \irrep{35}, \irrep{189}, \irrep{175} of 
\SU{6} and \irrep{1}, \irrep{24} and \irrep{75} of \SU{5} do not contribute 
chiral fermions and are disregarded here. Decomposing the \SU{N} two-column 
tableaux irreps to \SU{5} using 
\eqref{eqn:TwoColumnTableauxSU6toSU5Decomposition} we obtain an inhomogeneous 
system of linear equations for the multiplicities $m_i$:
\newcolumntype{x}{>{\raggedleft\hspace{0pt}}p{2ex}}
\begin{equation}
\left[
\begin{array}{@{}xxxxxx@{\;\:}|@{\;}x}
 1 & 1 & 1 & 0 & 0 & 0  & -3 \tabularnewline
 0 & 1 & 1 & 0 & 0 & 0  &  3 \tabularnewline
 1 & 1 & 0 & 0 & -1 & 0 &  0 \tabularnewline
 0 & -1 & 0 & 1 & 1 & 1 &  0 \tabularnewline
 0 & 0 & 1 & 1 & 0 & 1  &  0 \tabularnewline
 0 & 0 & 0 & 0 & 1 & 1  &  0 
\end{array}
\right]
\end{equation}
Since the coefficient matrix is quadratic and of full rank the system has the 
unique solution given by $m_1{\to}{-}6$, $m_2{\to}9$, $m_3{\to}{-}6$, 
$m_4{\to}9$, $m_5{\to}3$, $m_6{\to}{-}3$ which translates to 
\eqref{eqn:SU6Solution}.

In \SU{7} we have 9 complex, non-conjugated, two-column tableau irreps: 
$\irrep{28}$, $\irrep{112}$, $\irrep{140}$, $\irrep{196}$, $\irrep{210}$, 
$\irrep{224}$, $\irrep{490}$, $\irrep[1]{490}$ and $\irrep{588}$. The system of 
equations for the corresponding multiplicities $m_i$, with $i=1,\ldots,9$, is 
underdetermined leading to solution sets with three independent coefficients, 
$c_1$, $c_2$ and $c_3$:
\begin{equation}
	\begin{aligned}
	m_1 &\to c_1,\; m_2 \to c_2,\; m_3 \to c_1+2 c_3, \\
	m_4 &\to 3 c_1+2 c_2+2 c_3+6, \\
	m_5 &\to -20 c_1-8 c_2-19 c_3-51, \\
	m_6 &\to -16 c_1-7 c_2-16 c_3-36, \\
	m_7 &\to 20 c_1+8 c_2+20 c_3+51,    \\
	m_8 &\to -28 c_1-12 c_2-27 c_3-69,    \\
	m_9 &\to 13 c_1+6 c_2+12 c_3+30.
	\end{aligned}
\end{equation}

For individual solutions the independent coefficients ($c_j$s in general) take 
on positive and negative integer values. Simple solutions can be found by 
scanning through a limited range of integers for the $c_j$s, which we choose to 
be $c_j=-20,\ldots,20$, and we limit the total number of two-column tableau 
irreps to 20, i.e., $\sum_i |m_i|\leq20$. With these self imposed limitations, 
we find 9 solutions for \SU{7} displayed in a compact tabular form in terms of 
the multiplicities $m_i$ in Table~\ref{tab:TwoColumnTableauSU7Solutions},
\begin{table}
	\begin{tabular}{rrrrrrrrr}
	\toprule
		\irrep{28} & \irrep{112} & \irrep{140} & \irrep{196} & \irrep{210} & \irrep{224} & \irrep{490} & \irrep[1]{490} & \irrep{588} \\
	\midrule
		-2 &           1 &          -4 &           0 &           0 &           5 &          -1 &              2 &         -2  \\
		-1 &           1 &          -5 &           1 &          -1 &           5 &          -1 &              1 &         -1  \\
		 0 &           1 &          -6 &           2 &          -2 &           5 &          -1 &              0 &          0  \\
		-7 &           4 &          -1 &          -1 &           0 &           0 &           3 &             -2 &         -1  \\
		-6 &           4 &          -2 &           0 &          -1 &           0 &           3 &             -3 &          0  \\
		-3 &          -1 &          -1 &          -3 &          -2 &           3 &           3 &              0 &         -3  \\
		-2 &          -1 &          -2 &          -2 &          -3 &           3 &           3 &             -1 &         -2  \\
		-1 &          -1 &          -3 &          -1 &          -4 &           3 &           3 &             -2 &         -1  \\
		 0 &          -1 &          -4 &           0 &          -5 &           3 &           3 &             -3 &          0  \\
	\bottomrule
	\end{tabular}
	\caption{\label{tab:TwoColumnTableauSU7Solutions} Three family solutions for two-column tableau \SU{7} irreps}
\end{table} 
which translates to models with the following sets of \SU{7} fermion irreps:
\begin{scriptsize}
	\begin{equation}\label{eqn:SU7Solutions}
		\renewcommand*{\arraystretch}{1.5}
		\begin{array}{l}
			2(\irrepbar{28}){+}\irrep{112}{+}4(\irrepbar{140}){+}5(\irrep{224}){+}\irrepbar{490}{+}2(\irrep[1]{490}){+}2(\irrepbar{588})\\
			\irrepbar{28}{+}\irrep{112}{+}5(\irrepbar{140}){+}\irrep{196}{+}\irrepbar{210}{+}5(\irrep{224}){+}\irrepbar{490}{+}\irrep[1]{490}{+}\irrepbar{588}\\
			\irrep{112}{+}6(\irrepbar{140}){+}2(\irrep{196}){+}2(\irrepbar{210}){+}5(\irrep{224}){+}\irrepbar{490}\\
			7(\irrepbar{28}){+}4(\irrep{112}){+}\irrepbar{140}{+}\irrepbar{196}{+}3(\irrep{490}){+}2(\irrepbar[1]{490}){+}\irrepbar{588}\\
			6(\irrepbar{28}){+}4(\irrep{112}){+}2(\irrepbar{140}){+}\irrepbar{210}{+}3(\irrep{490}){+}3(\irrepbar[1]{490})\\
			3(\irrepbar{28}){+}\irrepbar{112}{+}\irrepbar{140}{+}3(\irrepbar{196}){+}2(\irrepbar{210}){+}3(\irrep{224}){+}3(\irrep{490}){+}3(\irrepbar{588})\\
			2(\irrepbar{28}){+}\irrepbar{112}{+}2(\irrepbar{140}){+}2(\irrepbar{196}){+}3(\irrepbar{210}){+}3(\irrep{224}){+}3(\irrep{490}){+}\irrepbar[1]{490}{+}2(\irrepbar{588})\\
			\irrepbar{28}{+}\irrepbar{112}{+}3(\irrepbar{140}){+}\irrepbar{196}{+}4(\irrepbar{210}){+}3(\irrep{224}){+}3(\irrep{490}){+}2(\irrepbar[1]{490}){+}\irrepbar{588}\\
			\irrepbar{112}{+}4(\irrepbar{140}){+}5(\irrepbar{210}){+}3(\irrep{224}){+}3(\irrep{490}){+}3(\irrepbar[1]{490})
		\end{array}
	\end{equation}
\end{scriptsize}

Moving on to \SU{8} we have 12 complex, non-conjugated, two-column tableau 
irreps: $\irrep{36}$, $\irrep{168}$, $\irrep{216}$, $\irrep{336}$, 
$\irrep{378}$, $\irrep{420}$, $\irrep{504}$, $\irrep{1008}$, $\irrep{1176}$, 
$\irrep{1344}$, $\irrep{1512}$ and $\irrep[1]{2352}$ and the system of equations 
leads to solution sets with six independent coefficients $c_j$:
\begin{equation}
	\begin{aligned}
		m_1   &\to c_1,\; m_2 \to c_2, \; m_3 \to c_3,\; m_4 \to c_4,\; m_5 \to c_5,\\
		m_6   &\to {-}33 c_1{-}48 c_2{-}30 c_3{-}28 c_4{-}105,                                              \\
		m_7   &\to 4 c_1{+}7 c_2{+}3 c_3{+}8 c_5{+}28 c_6,                                                \\
		m_8   &\to {-}33 c_1{-}47 c_2{-}30 c_3{-}27 c_4{+}2 c_5{+}c_6{-}108,                                    \\
		m_9   &\to 60 c_1{+}86 c_2{+}54 c_3{+}51 c_4{-}3 c_5{-}3 c_6{+}195,                                   \\
		m_{10}&\to 24 c_1{+}35 c_2{+}21 c_3{+}21 c_4{+}75,                                                \\
		m_{11}&\to 30 c_1{+}42 c_2{+}28 c_3{+}27 c_4{-}7 c_5{-}21 c_6{+}108,                                  \\
		m_{12}&\to {-}63 c_1{-}90 c_2{-}57 c_3{-}55 c_4{+}6 c_5{+}15 c_6{-}210.                                  
	\end{aligned}
\end{equation}
We find 11 solutions for a maximum of 20 two-column tableau irreps but with a 
smaller scan range for the six independent coefficients $c_j=-5,\ldots,5$, with 
$j=1,\ldots6$ as displayed in Table~\ref{tab:TwoColumnTableauSU8Solutions}.
\begin{table}[h]
	\begin{scriptsize}
	\begin{tabular}{rrrrrrrrrrrr}
	\toprule
		\irrep{36} & \irrep{168} & \irrep{216} & \irrep{336} & \irrep{378} & \irrep{420} & \irrep{504} & \irrep{1008} & \irrep{1176} & \irrep{1344} & \irrep{1512} & \irrep[1]{2352} \\
	\midrule
		-4 &  0 &  0 &  1 &  2 & -1 &  0 &  1 &  0 &  0 &  1 & -1 \\
		-2 &  1 & -2 & -1 &  1 &  1 &  1 &  0 & -1 & -1 &  0 &  1 \\
		 0 & -4 &  0 &  3 &  0 &  3 &  0 &  0 &  1 & -2 &  0 &  0 \\
		 1 & -1 & -3 &  0 &  2 &  0 &  4 &  0 &  1 &  1 & -2 &  0 \\
		-4 &  0 & -1 &  2 & -1 &  1 &  1 & -1 &  3 &  0 &  0 & -2 \\
		-1 &  1 & -4 &  0 &  1 &  0 & -1 &  0 &  2 &  2 &  1 & -3 \\
		-4 &  0 & -2 &  3 & -1 &  3 & -2 &  2 &  0 &  0 & -1 &  0 \\
		 0 & -1 & -1 & -1 &  1 &  1 & -2 & -2 &  1 & -2 &  4 & -2 \\
		-2 &  1 & -3 &  0 & -2 &  3 &  2 & -2 &  2 & -1 & -1 &  0 \\
		 2 & -3 &  0 & -1 &  2 &  1 &  3 & -2 &  0 & -3 &  1 &  1 \\
		 3 & -3 & -3 &  1 &  2 &  2 & -2 &  1 &  0 &  0 &  1 & -1 \\
	\bottomrule
	\end{tabular}
	\end{scriptsize}
	\caption{\label{tab:TwoColumnTableauSU8Solutions} Three family solutions for two-column tableau \SU{8} irreps}
\end{table}

Finally, for \SU{9} we obtain solution sets with 10 independent coefficients 
$c_j$ for the multiplicities of the 16 complex, non-conjugated, two-column 
tableau irreps $\irrep{45}$, $\irrep{240}$, $\irrep{315}$, $\irrep{540}$, 
$\irrep{630}$, $ \irrep{720}$, $\irrep{1008}$, $\irrep{1050}$, $\irrep{1890}$, 
$\irrep{2520}$, $ \irrep{2700}$, $\irrep{3402}$, $\irrep{3780}$, $\irrep{5292}$, 
$\irrep{6048}$ and $\irrep{7560}$: 
\begin{equation}
	\begin{aligned}
		m_1    \to &\  c_1,m_2\to c_2,m_3\to c_3,m_4\to c_4,m_5\to c_5,\\
		m_6    \to &\  2 c_1{+}3 c_2{+}2 c_3{+}6 c_6,
		m_7    \to     2 c_3{+}2 c_5{+}3 c_7,\\
		m_8    \to &\  3 c_1{+}3 c_3{+}3 c_5{+}4 c_8, 
		m_9    \to     c_9,\\
		m_{10} \to &\  31 c_1{+}33 c_2{+}8 c_3{+}43 c_5{+}44 c_6 {+}19 c_7\\
		           &\ {+}10 c_8{+}30 c_9{+}57 c_{10}{+}54,\\
		m_{11} \to &\  29 c_1{+}27 c_2{+}3 c_3{-}4 c_4{+}45 c_5{+}38 c_6\\
		           &\ {+}21 c_7{+}11 c_8{+}36 c_9{+}63 c_{10}{+}56,\\
		m_{12} \to &\ {-}263 c_1{-}270 c_2{-}58 c_3{+}20 c_4{-}378 c_5{-}372 c_6\\
		           &\ {-}178 c_7{-}86 c_8{-}291 c_9{-}518 c_{10}{-}483,\\
		m_{13} \to &\ {-}185 c_1{-}185 c_2{-}41 c_3{+}15 c_4{-}263 c_5{-}258 c_6\\
		           &\ {-}119 c_7{-}65 c_8{-}200 c_9{-}357 c_{10}{-}329,\\
		m_{14} \to &\ {-}773 c_1{-}790 c_2{-}167 c_3{+}60 c_4{-}1107 c_5{-}1092 c_6\\
		           &\ {-}514 c_7{-}256 c_8{-}851 c_9{-}1518 c_{10}{-}1411,\\
		m_{15} \to &\  485 c_1{+}495 c_2{+}103 c_3{-}40 c_4{+}698 c_5{+}685 c_6\\
		           &\ {+}327 c_7{+}160 c_8{+}540 c_9{+}960 c_{10}{+}892\\
		m_{16} \to &\ 220 c_1{+}224 c_2{+}51 c_3{-}15 c_4{+}310 c_5{+}310 c_6\\
		           &\ {+}140 c_7{+}75 c_8{+}234 c_9{+}420 c_{10}{+}390                     
	\end{aligned}
\end{equation}
and find 11 solutions for a maximum of 27 two-column tableau irreps and 
$c_j=-1,\ldots,1$, with $j=1,\ldots,10$ displayed in 
Table~\ref{tab:TwoColumnTableauSU9Solutions}.
\begin{table*}
	\begin{tabular}{rrrrrrrrrrrrrrrr}
	\toprule
	\irrep{45} & \irrep{240} & \irrep{315} & \irrep{540} & \irrep{630} & \irrep{720} & \irrep{1008} & \irrep{1050} & \irrep{1890} & \irrep{2520} & \irrep{2700} & \irrep{3402} & \irrep{3780} & \irrep{5292} & \irrep{6048} & \irrep{7560}\\
	\midrule
		 0 & -1 &  1 & -1 &  0 & -1 &  2 &  3 & -1 & -1 &  0 &  0 &  0 &  3 &  0 & -2 \\
		-1 &  1 & -1 &  0 &  0 & -1 & -2 & -2 &  0 &  1 & -1 &  0 &  4 &  1 & -1 & -2 \\
		 1 & -1 &  1 & -1 &  0 &  1 & -1 &  2 & -1 &  1 & -3 &  1 & -1 &  0 & -2 &  3 \\
		 1 & -1 &  1 & -1 & -1 &  1 & -3 &  3 &  0 & -2 & -1 &  2 & -3 &  0 &  0 &  2 \\
		 0 &  0 &  0 &  1 &  0 & -6 & -3 &  4 &  0 &  1 &  4 &  1 & -2 & -1 &  0 &  0 \\
		 0 &  0 &  0 &  1 &  1 & -6 & -1 &  3 & -1 &  4 &  2 &  0 &  0 & -1 & -2 &  1 \\
		 0 &  0 & -1 &  1 &  0 & -2 & -5 & -3 &  1 &  0 &  1 &  0 &  3 & -3 &  2 & -2 \\
		 0 &  1 & -1 &  0 & -1 &  1 & -7 & -2 &  1 &  0 & -2 &  2 &  1 & -2 & -1 &  2 \\
		 1 &  0 & -1 &  0 &  1 & -6 & -3 &  3 &  0 &  0 &  5 &  2 & -2 &  0 &  0 & -1 \\
		 0 & -1 &  1 & -1 & -1 & -1 &  0 &  4 &  0 & -4 &  2 &  1 & -2 &  3 &  2 & -3 \\
		 0 &  0 & -1 &  1 &  1 & -2 & -3 & -4 &  0 &  3 & -1 & -1 &  5 & -3 &  0 & -1 \\	
	\bottomrule
	\end{tabular}
	\caption{\label{tab:TwoColumnTableauSU9Solutions} Three family solutions for two-column tableau \SU{9} irreps}
\end{table*}

We do not need to limit ourselves to two-column tableaux. For instance we can 
search for three column sets that are exotic free, anomaly free and have three 
families. Here we conclude with the two- and three-column \SU{6} case, where we 
find solution sets with six independent coefficients $c_j$:
\begin{equation}
	\begin{aligned}
		m_1    & \to c_1, m_2\to {-}c_1{-}6, m_3\to c_2,m_4\to c_3,\\
		m_5    & \to c_4, m_6 \to c_5, m_7\to c_6,m_8\to 6{-}c_2,\\
		m_9    & \to c_2{+}c_6{-}9,m_{10} \to {-}c_6, m_{11} \to {-}c_3{-}c_4{-}c_6{+}3,     \\
		m_{12} & \to {-}c_3{-}6,m_{13}  \to {-}c_1{-}c_5{-}c_6{-}3,m_{14}  \to c_2{-}c_5{-}6,\\
		m_{15} & \to c_1{-}c_3{+}c_6, m_{16}  \to 9{-}c_4,\\
		m_{17} & \to {-}c_1{+}c_3{-}c_5{-}c_6{+}3, m_{18}  \to {-}c_3{-}6,\\
		m_{19} & \to c_2{-}9, m_{20} \to c_1{+}c_4{+}c_6{-}3,\\
		m_{21} & \to {-}c_1{-}c_4{-}c_6{+}3, m_{22} \to c_1{-}c_2{+}c_4{+}c_5{+}c_6{+}3.
	\end{aligned}
\end{equation}

With a maximum of 62 two- and three-column tableau irreps and $c_j=-2,\ldots,2$, 
with $j=1,\ldots6$ we find 17 solutions displayed in 
Table~\ref{tab:ThreeColumnTableauSU6Solutions}.

\begin{table*}
	\begin{tabular}{rrrrrrrrrrrrrrrrrrrrrr}
	\toprule
	\irrep{21} & \irrep{56} & \irrep{70} & \irrep{84} & \irrep{105} & \irrep[1]{105} & \irrep{120} & 
	\irrep{210} & \irrep[1]{210} & \irrep{280} & \irrep{336} & \irrep{384} & \irrep{420} & \irrep{490} & 
	\irrep{560} & \irrep{840} & \irrep[1]{840} & \irrep{896} & \irrep{1050} & \irrep{1176} & \irrep[1]{1176} & \irrep{1470}\\
	\midrule
		-2 & -4 & 2 & -2 & 2 & -2 & 1 & 4 & -6 & -1 & 2 & -4 & 0 & -2 & 1 & 7 & 4 & -4 & -7 & -2 & 2 & 0 \\
		-2 & -4 & 2 & -2 & 2 & -2 & 2 & 4 & -5 & -2 & 1 & -4 & -1 & -2 & 2 & 7 & 3 & -4 & -7 & -1 & 1 & 1 \\%
		-2 & -4 & 2 & -1 & 2 & -2 & 1 & 4 & -6 & -1 & 1 & -5 & 0 & -2 & 0 & 7 & 5 & -5 & -7 & -2 & 2 & 0 \\
		-2 & -4 & 2 & -1 & 2 & -2 & 2 & 4 & -5 & -2 & 0 & -5 & -1 & -2 & 1 & 7 & 4 & -5 & -7 & -1 & 1 & 1 \\
		-1 & -5 & 2 & -2 & 2 & -2 & 0 & 4 & -7 & 0 & 3 & -4 & 0 & -2 & 1 & 7 & 4 & -4 & -7 & -2 & 2 & 0 \\
		-1 & -5 & 2 & -2 & 2 & -2 & 1 & 4 & -6 & -1 & 2 & -4 & -1 & -2 & 2 & 7 & 3 & -4 & -7 & -1 & 1 & 1 \\%
		-1 & -5 & 2 & -2 & 2 & -2 & 2 & 4 & -5 & -2 & 1 & -4 & -2 & -2 & 3 & 7 & 2 & -4 & -7 & 0 & 0 & 2 \\
		-1 & -5 & 2 & -1 & 2 & -2 & 0 & 4 & -7 & 0 & 2 & -5 & 0 & -2 & 0 & 7 & 5 & -5 & -7 & -2 & 2 & 0 \\
		-1 & -5 & 2 & -1 & 2 & -2 & 1 & 4 & -6 & -1 & 1 & -5 & -1 & -2 & 1 & 7 & 4 & -5 & -7 & -1 & 1 & 1 \\%
		-1 & -5 & 2 & -1 & 2 & -2 & 2 & 4 & -5 & -2 & 0 & -5 & -2 & -2 & 2 & 7 & 3 & -5 & -7 & 0 & 0 & 2 \\
		-1 & -5 & 2 & 0 & 2 & -2 & 1 & 4 & -6 & -1 & 0 & -6 & -1 & -2 & 0 & 7 & 5 & -6 & -7 & -1 & 1 & 1 \\
		0 & -6 & 2 & -2 & 2 & -2 & 0 & 4 & -7 & 0 & 3 & -4 & -1 & -2 & 2 & 7 & 3 & -4 & -7 & -1 & 1 & 1 \\
		0 & -6 & 2 & -2 & 2 & -2 & 1 & 4 & -6 & -1 & 2 & -4 & -2 & -2 & 3 & 7 & 2 & -4 & -7 & 0 & 0 & 2 \\
		0 & -6 & 2 & -1 & 2 & -2 & 0 & 4 & -7 & 0 & 2 & -5 & -1 & -2 & 1 & 7 & 4 & -5 & -7 & -1 & 1 & 1 \\
		0 & -6 & 2 & -1 & 2 & -2 & 1 & 4 & -6 & -1 & 1 & -5 & -2 & -2 & 2 & 7 & 3 & -5 & -7 & 0 & 0 & 2 \\
		0 & -6 & 2 & 0 & 2 & -2 & 0 & 4 & -7 & 0 & 1 & -6 & -1 & -2 & 0 & 7 & 5 & -6 & -7 & -1 & 1 & 1 \\
		0 & -6 & 2 & 0 & 2 & -2 & 1 & 4 & -6 & -1 & 0 & -6 & -2 & -2 & 1 & 7 & 4 & -6 & -7 & 0 & 0 & 2 \\
	\bottomrule
	\end{tabular}
	\caption{\label{tab:ThreeColumnTableauSU6Solutions} Three family solutions for two- and three-column tableau \SU{6} irreps}
\end{table*}

Other cases are also easily explored. For instance we could consider 
combinations of one and two-columns tableau, or just three column tableau, etc. 
We could also redo the above analysis for four families. Alternatively, we could 
study anomaly free three family models with a specific set of exotics. All these 
possibilities as well as other types of model scans (See e.g., 
\cite{Albright:2012zt}.) can be easily handled with LieART~\cite{Feger:2012bs}.

\section{\boldmath An \SU{6} example} 
\newcolumntype{y}{>{\raggedleft\hspace{0pt}}p{3.8ex}}
\newcolumntype{z}{>{\raggedleft\hspace{0pt}}p{2.5ex}}
\begin{table*}[t!]
	\begin{scriptsize}
	\begin{tabularx}{\textwidth}{p{3.5em}@{\!}lX}
	\toprule
	\boldmath\textbf{$\SU{N}$} & \;\,\textbf{Equation system} & \textbf{One-family model solutions}\\
	\midrule
	\SU{7} & \setlength{\arraycolsep}{2pt}
			$\left[
			\begin{array}{@{}zzzzzzzzz@{\;\,}|@{\:}z}
			2 & 4 & 3 & -2 & -2 & 2 & -1 & 0 & 1 & -1 \tabularnewline
			 0 & 2 & 2 & -1 & -2 & 2 & -1 & 0 & 1 &  1 \tabularnewline
			 1 & 2 & 0 & -3 & -1 & 0 & -2 & -1 & 0 &  0 \tabularnewline
			 0 & -1 & 0 & 2 & 2 & 1 & 4 & 2 & 2 &  0 \tabularnewline
			 0 & 0 & 1 & 0 & 1 & 2 & 2 & 1 & 3 &  0 \tabularnewline
			 0 & 0 & 0 & 1 & 0 & 0 & 2 & 2 & 1 &  0 
			\end{array}\right]$
		 &
		$\begin{array}{X}   
			4(\irrepbar{28})+3(\irrep{112})+\irrep{210}+\irrepbar{224}+\irrep{490}+\irrepbar[1]{490}\\
			2(\irrepbar{112})+2(\irrepbar{196})+\irrepbar{210}+2(\irrep{224})+\irrep{490}+\irrep[1]{490}+2(\irrepbar{588})\\               
			\irrep{28}+2(\irrepbar{112})+\irrepbar{140}+\irrepbar{196}+2(\irrepbar{210})+2(\irrep{224})+\irrep{490}+\irrepbar{588}\\
			3(\irrepbar{28})+3(\irrep{112})+\irrepbar{140}+\irrep{196}+\irrepbar{224}+\irrep{490}+2(\irrepbar[1]{490})+\irrep{588}\\
			2(\irrep{28})+2(\irrepbar{112})+2(\irrepbar{140})+3(\irrepbar{210})+2(\irrep{224})+\irrep{490}+\irrepbar[1]{490}\\
			5(\irrepbar{28})+3(\irrep{112})+\irrep{140}+\irrepbar{196}+2(\irrep{210})+\irrepbar{224}+\irrep{490}+\irrepbar{588}\\
	       \end{array}$\\ \midrule                                                                                                                 
	\SU{8} & \setlength{\arraycolsep}{2pt}
			$\left[
			\begin{array}{@{}zzzzzzzzzzzz@{\;\,}|@{\:}z}
			3 & 9 & 6 & 8 & 9 & 8 & 0 & -9 & -3 & 6 & 0 & 0 & -1 \tabularnewline
			0 & 3 & 3 & 3 & 6 & 6 & 0 & -6 & -2 & 5 & 0 & 0 &  1 \tabularnewline
			1 & 3 & 0 & 6 & 3 & 0 & -1 & -8 & -6 & 0 & -3 & -3 &  0 \tabularnewline
			0 & -1 & 0 & -3 & -3 & 1 & 3 & 9 & 8 & 3 & 9 & 8 &  0 \tabularnewline
			0 & 0 & 1 & 0 & -1 & 3 & 3 & 3 & 3 & 6 & 8 & 6 &  0 \tabularnewline
			0 & 0 & 0 & -1 & 0 & 0 & 0 & 3 & 5 & 1 & 3 & 5&  0 
			\end{array}\right]$
		 & 
		$\begin{array}{X}   
			\irrepbar{36}+\irrepbar{216}+\irrep{336}+\irrep{378}+\irrep{504}+2(\irrep{1008})+\irrepbar{1176}+\irrep{1344}+2(\irrepbar{1512})+\irrep[1]{2352}\\
			\irrep{36}+2(\irrepbar{168})+\irrep{336}+\irrep{378}+2(\irrepbar{504})+\irrep{1176}+2(\irrep{1512})+2(\irrepbar[1]{2352})\\
			4(\irrepbar{36})+2(\irrep{168})+\irrep{216}+\irrepbar{336}+\irrepbar{420}+\irrep{504}+\irrepbar{1008}+\irrepbar{1344}+\irrep{1512}\\
			3(\irrep{36})+\irrepbar{168}+2(\irrepbar{216})+\irrepbar{336}+2(\irrep{420})+\irrepbar{504}+\irrepbar{1008}+\irrepbar{1344}+\irrep{1512}\\
			\irrepbar{36}+\irrep{378}+2(\irrepbar{420})+4(\irrep{504})+\irrepbar{1008}+2(\irrep{1176})+\irrep{1344}+\irrepbar{1512}+\irrepbar[1]{2352}\\
			2(\irrepbar{168})+2(\irrep{216})+\irrep{378}+\irrep{420}+2(\irrepbar{1176})+3(\irrepbar{1344})+\irrep{1512}+2(\irrep[1]{2352})\\
	       \end{array}$\\ \midrule
	\SU{9} & \setlength{\arraycolsep}{1pt}
			$\left[
			\begin{array}{@{\!\!\!}yyyyyyyyyyyyyyyy@{\;}|@{\!}y}
			 4 & 16 & 10 & 20 & 24 & 20 & -12 & 11 & -35 & -20 & 20 & -18 & 14 & -3 & -12 & 6 & -1 \tabularnewline
			 0 & 4 & 4 & 6 & 12 & 12 & -8 & 8 & -18 & -11 & 14 & -12 & 11 & -2 & -8 & 5 &  1 \tabularnewline
			 1 & 4 & 0 & 10 & 6 & 0 & -4 & -1 & -20 & -20 & 0 & -15 & -4 & -9 & -20 & -6 &  0 \tabularnewline
			 0 & -1 & 0 & -4 & -4 & 1 & 6 & 4 & 16 & 20 & 4 & 24 & 16 & 16 & 34 & 20 &  0 \tabularnewline
			 0 & 0 & 1 & 0 & -1 & 4 & 4 & 6 & 4 & 6 & 10 & 15 & 20 & 9 & 20 & 20 &  0 \tabularnewline
			 0 & 0 & 0 & -1 & 0 & 0 & 0 & 0 & 4 & 9 & 1 & 6 & 4 & 10 & 16 & 9 &  0 
			\end{array}\right]$
		 & 
		$\begin{array}{X}   
			\irrep{45}+\irrepbar{240}+\irrepbar{315}+\irrep{720}+\irrepbar{1890}+\irrep{2520}+\irrep{3402}+\irrepbar{3780}+\irrepbar{6048}+\irrep{7560}\\
			\irrepbar{240}+\irrep{540}+\irrepbar{630}+\irrep{720}+\irrep{1050}+\irrepbar{1890}+2(\irrep{2520})+\irrepbar{2700}+\irrepbar{3780}\\
			\qquad+\irrepbar{5292}+\irrepbar{6048}+2(\irrep{7560})\\
			\irrepbar{45}+\irrepbar{315}+\irrepbar{540}+\irrep{630}+2(\irrep{1008})+\irrep{1050}+2(\irrepbar{2520})+2(\irrep{2700})\\
			\qquad+2(\irrepbar{3780})+2(\irrep{6048})+\irrepbar{7560}\\
			\irrepbar{315}+\irrepbar{540}+\irrep{630}+2(\irrep{720})+\irrepbar{1008}+\irrepbar{2700}+\irrep{3402}+3(\irrepbar{3780})\\
			\qquad+3(\irrepbar{5292})+4(\irrep{7560})\\
	      \end{array}$\\
	\bottomrule
	\end{tabularx}
	\end{scriptsize}
	\caption{\label{tab:TwoColumnTableauOneFamilySolutions} One family equation systems and solutions for two-column tableau irreps}
\end{table*}%
Besides the three family exotic models discussed above, we should also display 
the simplest of all models found to date that starts with any number of 
multicolumn tableaux plus some single column tableaux that has three families. 
Since we  already have three families in \SU{6} for our two-column example in 
\eqref{eqn:SU6Solution} and as all coefficients are a multiple of 3, we must 
have one family if we divide all coefficients by three. Hence we can add the 
single column irreps $4(\irrepbar{6}) + 2(\irrep{15})$ to this set to get a 
three family model
\begin{equation}
	\begin{split}
		2(\irrepbar{21}) + 3(\irrep{70}) + 2(\irrepbar{84}) + 3(\irrep{105}) + \irrep[1]{105} + \irrepbar{210}\\
		+4(\irrepbar{6}) + 2(\irrep{15}) = 3(\irrepbar{5} + \irrep{10}) + \text{real}
	\end{split}
\end{equation}
It seems most natural to let the two lightest families be in the $4(\irrepbar{6}) + 
2(\irrep{15})$ and the third family to be the ``exotic'' family.

While this example may not be simple enough to be a useful physical model, it is 
still instructive to examine it further. For instance, if we break the symmetry 
along the path $\SU{6} \rightarrow \SU{5} \times \U{1}'$ then as long as the 
extra $\U{1}'$ is unbroken, some of the \SU{5} conjugate pair exotics (as well 
as some $( \irrep{5}+ \irrepbar{5})$ and $(\irrep{10}+\irrepbar{10})$ pairs) 
stay light, as long as their $\U{1}'$ charges are imbalanced. This remains true 
even if we break to $ \SU{3} \times \SU{2} \times \U{1} \times \U{1}' $, but 
when we break to the standard model $\SU{3} \times \SU{2} \times \U{1}$ gauge 
group all the exotics can finally acquire mass.

If we were to keep $\U{1}'$ unbroken until $\sim 1$ TeV, then we would predict 
very many light (TeV scale) exotic fermions. Since keeping the extra $\U{1}'$ 
does not directly lead to proton decay it is probably allowed to be unbroken 
down nearly to the electroweak scale. However, since this model leads to so many 
exotics, a low energy $\U{1}'$ would undoubtedly upset the renormalization group 
running and spoil unification. So we conjecture that the best we can do is bring 
the $\U{1}'$ scale down a few orders of magnitude from the GUT scale. This model 
is by no means compelling, but it is still interesting, as it is the first 
example of a type of model with exotic fermions that can exist well below the 
GUT scale. As we noted above, better would be a model with only a few light 
exotics and a low energy $\U{1}'$ where the exotics could even be within reach 
of the LHC.

Other one-family exotic models can be found directy with our algorithm by 
requiring the decomposition to only one set of $\irrepbar{5}+\irrep{10}$ and all 
other fermions to be vector-like. In 
Table~\ref{tab:TwoColumnTableauOneFamilySolutions} we list the one-family model 
equation systems and some solutions for two-column tableaux for \SU{7}, \SU{8} 
and \SU{9}. We have three column examples but they are complicated and not very 
enlightening, so we have chosen not to display them.

\section{Conclusions} 

We have explored \SU{N} gauge theory examples that start with mixed tensor 
fermonic irreps that none the less have only three standard families of chiral 
fermions at the \SU{5} level. These results have been obtained with LieART, 
which is a programmable group theory software package capable of handling such 
complicated tasks. If we relax the constraint of starting with 20 irreps and a 
limited scan range for the independent coefficients, then there is an 
arbitrarily large class of models that start with chiral exotic fermions (i.e., 
fermions in multicolumn tableaux) at the \SU{N} level, but where there are only 
standard chiral families at the \SU{5} and SM level. While so far none of these 
models are particularly compelling, the results do demonstrate a new avenue for 
model building. It is conceivable that a model like one of these could describe 
the UV completion of the SM. Although at present we do not have an example, that 
such models could arise remains a logical possibility. We plan to search for 
such models.

\section*{Note Added in Proof}
The chirality and fermionic particle content of the SM coming from grand unified 
theories has been investigated from a somewhat different point of view in 
\cite{Fonseca:2015aoa}. Where results overlap with our work they agree.
\vfill

\acknowledgments
Most of the work was performed while RPF was affiliated with the Department of 
Physics and Astronomy, Vanderbilt University, Nashville, and his work was 
supported by a fellowship within the Postdoc-Programme of the German Academic 
Exchange Service (DAAD) and by US DoE grant DE-FG05-85ER40226. The work of TWK 
was supported by US DoE grants DE-FG05-85ER40226 and DE-SC0010504.
\vfill

\pagebreak
\bibliography{references}

\end{document}